\documentclass[slac_one]{revtex4}
\usepackage{graphicx}
\usepackage{fancyhdr}
\begin{document}

\title{The study of heavy flavors via non-photonic electrons in STAR} 

\author{J. Bielcik for STAR collaboration}

\affiliation{Faculty of Nuclear Sciences and Physical Engineering, Czech Technical University in Prague, Prague, Czech Republic}
\affiliation{Nuclear Physics Institute of the ASCR, v.v.i, Rez, Czech Republic}

\begin{abstract}
The strong suppression of hadrons with high transverse momenta in central
Au+Au collisions observed at RHIC is generally interpreted as a consequence
of energy loss of energetic partons in the hot and dense matter before
fragmenting. The study of heavy quark production tests our understanding of
energy loss mechanisms.

Heavy quarks were expected to lose less energy in the medium than light
quarks and gluons due to the suppression of small-angle gluon radiation.
However, the high transverse momentum non-photonic electron spectra, which
are dominated by semi-leptonic decays of heavy quarks, show a strong
suppression in central Au+Au collisions as well. Current theoretical models
do not satisfactory explain this observation.

Preliminary non-photonic electron spectra are being extracted for Cu+Cu
collisions at $\sqrt{s_{NN}}$ = 200 GeV. They are compared to the non-photonic
electron yields from p+p and Au+Au collisions at the same collision energy.
This provides a direct experimental test of the dependence of the
non-photonic electron yield on collision system size.
\end{abstract}

\maketitle

\thispagestyle{fancy}

\section{INTRODUCTION} 
    During last several years collisions of the heavy ions at RHIC 
 have been used to study the nuclear matter under the conditions 
of the high temperature and energy density \cite{star_white}. 
The observation of the  suppression of inclusive hadrons  with high transverse momentum ($p_{T}$) in 
central Au+Au collisions is generally interpreted as a consequence of energy loss of energetic light partons 
in nuclear matter. The understanding of energy loss mechanism is an essential part when extracting 
the information about the properties of nuclear matter created in the collision.
    Heavy quarks, such as charm and bottom, are created in the initial phase of the collision in 
hard scattering~\cite{Lin:1995pk} and they are therefore sensitive to all later phases  of 
the system evolution. All partons lose their energy  while traversing the medium via gluon bremsstrahlung.
However there is a mass dependent suppression of this radiation under angles smaller than ratio of quark mass and quark energy ~\cite{Dokshitzer:2001zm}. This effect is expected to significantly reduce the energy loss of heavy quarks when compared with light partons. Although there is an elastic energy loss, that is expected to be an important contribution in the case of heavy quarks~\cite{mustafa,djordjevic}.  
    The STAR experiment has studied heavy quarks in several different channels: directly in D meson reconstruction using hadronic channels, indirectly using  decays to electrons and muons. The spectrum of  electrons measured by STAR is after subtracting the contribution from $\gamma$ conversions and Dalitz decays (mainly $\pi_{0}$ and $\eta$)  dominated by non-photonic electrons (both electrons and positrons are ment in the following) from semileptonic decays of D and B mesons. Combined information from all three channels
is used to extract the inclusive charm production cross section. This has been performed in several collision systems at $\sqrt{s_{NN}}$~=~200~GeV: d+Au~\cite{STARcharmpaper}, Cu+Cu~\cite{baumgart} and Au+Au~\cite{charm:2008hja} collisions.   The extracted values of the inclusive charm cross section show a binary scaling, which supports the idea that the charm is produced in hard scattering.  The measured charm cross section is within the large uncertainties consistent with FONLL calculations~\cite{charm:2008hja,ramona2}. 

\section{PREVIOUS MEASUREMENT IN AU+AU COLLISIONS}
    STAR has previously reported the yields of non-photonic electrons with high-{$p_{T}$} in p+p, d+Au and Au+Au~\cite{Abelev:2007} at $\sqrt{s_{NN}}$~=~200~GeV.  When studying the effect of created nuclear matter on particle yields, the  nuclear modification factor ($R_{AA}$) is commonly used. $R_{AA}$ of non-photonic electrons is the ratio of the yield measured in Au+Au to the yield from p+p collisions  - scaled by the mean number of binary collisions in the given collision system. If $R_{AA}$ would be a unity at high $p_{T}$, then the nucleus-nucleus collision can be viewed as an incoherent superposition of nucleon-nucleon collisions. This would  be the case if no hot and dense nuclear medium would be produced or it would not modify the particle production. 
Deviations from the unity measure the effects of the nuclear matter or  the quark-gluon plasma on particle yields.  A strong suppression in central Au+Au collisions as compared to p+p collisions was observed.  For electrons with 
$p_{T}>$~5~GeV/$c$ the value of nuclear modification factor $R_{AA}$~=0.2-0.3 was measured.
This suppression is as strong as the suppression of the inclusive 
light-flavor hadron yields~\cite{Adams:2003kv,Abelev:2007ra,Abelev:2006jr}. Several models attempted to describe the non-photonic 
suppression in central Au+Au collisions~\cite{Armesto:2005mz,Djordjevic:2005,Wicks:2005gt,vanHees:2005wb,adil2007}. In general, the models overpredict the observed suppression 
of non-photonic electrons when the parameters of these models (e.g. $dN_{g}/dy$)  are constrained by
 the suppresion of light hadrons. Even in the case of considering  both radiative energy loss and elastic energy loss of heavy quarks in the nuclear 
medium, the predicted $R_{AA}$ is too high at high-$p_{T}$. Only if a contribution of
bottom quarks is not taken into account, then the measured suppression is described well.

\begin{figure*}[t]
\centering
\includegraphics[width=135mm]{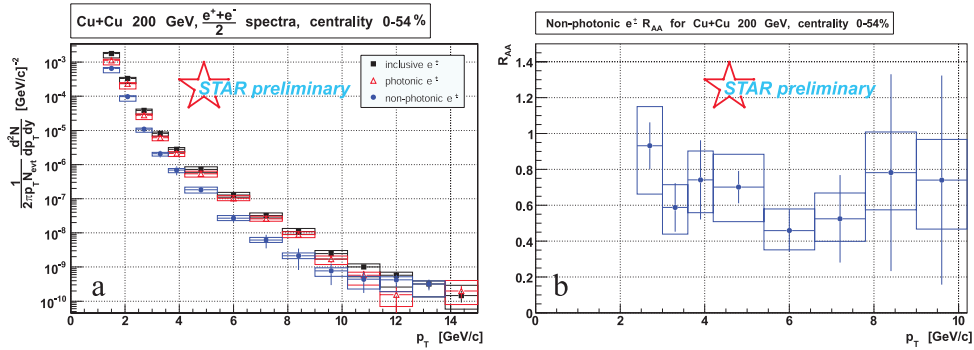}
\caption{Results from Cu+Cu collisions at $\sqrt{s_{NN}}$~=~200~GeV in the 0-54$\%$ centrality bin: a) inclusive, photonic, non-photonic electron yields with statistical errors (lines) and systematic errors (boxes). b) nuclear modification factor for non-photonic electrons.} \label{cucu200}
\end{figure*}

\section{NON-PHOTONIC ELECTRONS IN CU+CU COLLISIONS}
   In 2005 Cu+Cu collisions at $\sqrt{s_{NN}}$~=~200~GeV were recorded at RHIC. The results presented here represent  10 million minimum bias events  and 1.9 million high-tower triggered events. The high-tower trigger selects
the events where at least in one tower (segment) of the Barrel Electromagnetic Calorimeter (BEMC) was deposited energy above a given threshold (approximately 3.75~GeV/$c$ for the data presented in this paper).
   The electrons are identified by combining the information about the energy loss $dE/dx$ from the Time Projection Chamber (TPC), the deposited energy in BEMC and the shape of the electromagnetic shower reconstructed in the Shower Maximum Detector (SMD). The SMD is a subdetector of the BEMC located about five radiation legths inside. The total electron reconstruction efficiency was extracted from embeding simulated electrons in real Cu+Cu events. The details of the analysis steps can be found in~\cite{knospe}. Most of the photonic background is produced in low invariant mass (below 150~MeV/$c^{2}$) electron-positron pairs. These pairs are reconstructed and the random combinations are subtracted. Then the yield of the photonic electrons is subtracted from the inclusive electron continuum. An efficiency of this procedure (70-80$\%$) is a function of $p_{T}$) and it was extracted from embedingthe simulated $\pi^{0}$ in real Cu+Cu events.
  In Fig.1a the spectrum of reconstructed inclusive, photonic and non-photonic electrons for Cu+Cu collisions
 in the centrality 0-54$\%$ ($\langle N_{bin}\rangle$~=~82) with $p_{T}$ up to 15~GeV/$c$ is shown. The spectra extends well in the region above 5~GeV/$c$, where the contribution from electrons from B decays is important. 
  In Fig.1b the nuclear modification factor  $R_{AA}$ is presented. For $p_{T}>$~3~GeV/$c$ the value is
$R_{AA}$~=~$0.62+/-0.10(stat.)^{+0.14}_{-0.16}(sys.)$.

\section{DISCUSSION AND CONCLUSIONS}
  The suppression of non-photonic electrons measured in Cu+Cu collisions can be compared to the suppression 
of hadrons in the same system and centrality. The STAR experiment has measured the production of charged pions in Cu+Cu collisions~\cite{holis}. The observed suppression of charged pions with $p_{T}>$~2.5~GeV/$c$ in the close centrality bin 20-40$\%$  is 0.6-0.8. The measured value of  $R_{AA}$ is also consistent with the suppression observed for similar $\langle N_{bin}\rangle$ in Au+Au collisions at the same energy~\cite{suaide}. 
   In summary, the non-photonic electrons in Cu+Cu collisions at $\sqrt{s_{NN}}$~=~200~GeV were measured by STAR.
The observed nuclear modification factor $R_{AA}$~=~$0.62+/-0.10(stat.)^{+0.14}_{-0.16}(sys.)$ 
is consistent with the suppression of charged pions and the previous measurement of non-photonic electrons in Au+Au collisions.

\begin{acknowledgments}
This work was supported in part by the IRP AV0Z10480505,
by GACR grant 202/07/0079 and by grant LC07048 of the
Ministry of Education of the Czech Republic.
\end{acknowledgments}


\end{document}